\input harvmac
\overfullrule=0pt
\parindent 25pt
\tolerance=10000


\input epsf

\def\frac#1#2{{#1\over #2}}
\def\lp{\lambda'}
\def\l{{\lambda}}
\def\CO{{\cal O}}
\def\<{\langle}
\def\>{\rangle}
\def\H{{\cal H}}
\def\CN{{\cal N}}

\lref\GursoyYY{
U.~G\"ursoy,
``Vector operators in the BMN correspondence,''
arXiv:hep-th/0208041.
}

\lref\VerlindeIG{
H.~Verlinde,
``Bits, matrices and 1/N,''
arXiv:hep-th/0206059.
}

\lref\alg{
M.~Hatsuda, K.~Kamimura and M.~Sakaguchi,
``From super-AdS(5) x S**5 algebra to super-pp-wave algebra,''
Nucl.\ Phys.\ B {\bf 632}, 114 (2002)
[arXiv:hep-th/0202190].
}

\lref\sz{
A.~Santambrogio and D.~Zanon,
arXiv:hep-th/0206079.
}

\lref\GreenTC{
M.~B.~Green and J.~H.~Schwarz,
``Superstring Interactions,''
Nucl.\ Phys.\ B {\bf 218}, 43 (1983).
}

\lref\GreensiteHM{
J.~Greensite and F.~R.~Klinkhamer,
``Superstring Amplitudes And Contact Interactions,''
Nucl.\ Phys.\ B {\bf 304}, 108 (1988).
}

\lref\BerensteinSA{
D.~Berenstein and H.~Nastase,
``On lightcone string field theory from super Yang-Mills and holography,''
arXiv:hep-th/0205048.
}

\lref\bmn{
D.~Berenstein, J.~M.~Maldacena and H.~Nastase,
``Strings in flat space and pp waves from N = 4 super Yang Mills,''
JHEP {\bf 0204}, 013 (2002)
[arXiv:hep-th/0202021].
}

\lref\G{
N.~R.~Constable, D.~Z.~Freedman, M.~Headrick, S.~Minwalla, L.~Motl, A.~Postnikov and W.~Skiba,
``PP-wave string interactions from perturbative Yang-Mills theory,''
JHEP {\bf 0207}, 017 (2002)
[arXiv:hep-th/0205089].
}

\lref\germtwo{
N.~Beisert, C.~Kristjansen, J.~Plefka, G.~W.~Semenoff and M.~Staudacher,
``BMN Correlators and Operator Mixing in N=4 Super Yang-Mills Theory,''
arXiv:hep-th/0208178.
}

\lref\germone{
C.~Kristjansen, J.~Plefka, G.~W.~Semenoff and M.~Staudacher,
``A new double-scaling limit of N = 4 super Yang-Mills theory and PP-wave  strings,''
arXiv:hep-th/0205033.
}

\lref\hull{
C.~M.~Hull,
``Killing Spinors And Exact Plane Wave Solutions Of Extended Supergravity,''
Phys.\ Rev.\ D {\bf 30}, 334 (1984).
M.~Blau, J.~Figueroa-O'Farrill, C.~Hull and G.~Papadopoulos,
``A new maximally supersymmetric background of IIB superstring theory,''
JHEP {\bf 0201}, 047 (2002)
[arXiv:hep-th/0110242].
;M.~Blau, J.~Figueroa-O'Farrill, C.~Hull and G.~Papadopoulos,
``Penrose limits and maximal supersymmetry,''
Class.\ Quant.\ Grav.\  {\bf 19}, L87 (2002)
[arXiv:hep-th/0201081].

}

\lref\metsaev{
R.~R.~Metsaev,
``Type IIB Green-Schwarz superstring in plane wave Ramond-Ramond  background,''
Nucl.\ Phys.\ B {\bf 625}, 70 (2002)
[arXiv:hep-th/0112044].
R.~R.~Metsaev and A.~A.~Tseytlin,
``Exactly solvable model of superstring in plane wave Ramond-Ramond  background,''
Phys.\ Rev.\ D {\bf 65}, 126004 (2002)
[arXiv:hep-th/0202109].
}

\lref\svt{
M.~Spradlin and A.~Volovich,
``Superstring interactions in a pp-wave background II,''
arXiv:hep-th/0206073.
}

\lref\huang{
M.~x.~Huang,
``String interactions in pp-wave from N = 4 super Yang Mills,''
arXiv:hep-th/0206248.
}

\lref\lee{
P.~Lee, S.~Moriyama and J.~w.~Park,
``Cubic interactions in pp-wave light cone string field theory,''
arXiv:hep-th/0206065.
}

\lref\park{
Y.~j.~Kiem, Y.~b.~Kim, S.~m.~Lee and J.~m.~Park,
``pp-wave / Yang-Mills correspondence: An explicit check,''
arXiv:hep-th/0205279.
}

\lref\svo{
M.~Spradlin and A.~Volovich,
``Superstring interactions in a pp-wave background,''
arXiv:hep-th/0204146.
}

\lref\bianchi{
M.~Bianchi, B.~Eden, G.~Rossi and Y.~S.~Stanev,
``On operator mixing in N = 4 SYM,''
arXiv:hep-th/0205321.
}

\lref\huangtwo{
M.~x.~Huang,
``Three point functions of N = 4 super Yang Mills from light cone string  field theory in pp-wave,''
Phys.\ Lett.\ B {\bf 542}, 255 (2002)
[arXiv:hep-th/0205311].
}

\lref\chu{
C.~S.~Chu, V.~V.~Khoze and G.~Travaglini,
``Three-point functions in N = 4 Yang-Mills theory and pp-waves,''
JHEP {\bf 0206}, 011 (2002)
[arXiv:hep-th/0206005].
}

\lref\diffreg{
D.~Z.~Freedman, K.~Johnson and J.~I.~Latorre,
``Differential regularization and renormalization: A New method of calculation in quantum 
field theory,''
Nucl.\ Phys.\ B {\bf 371}, 353 (1992).
}

\lref\chutoo{
C.~S.~Chu, V.~V.~Khoze and G.~Travaglini,
``pp-wave string interactions from n-point correlators of BMN operators,''
arXiv:hep-th/0206167.
}

\lref\gross{
D.~J.~Gross, A.~Mikhailov and R.~ Roiban. 
``A calculation of the plane wave string Hamiltonian from N=4 super-Yang-Milss theory,''
archiv:hep-th/0208231}

\Title
{\vbox{
 \baselineskip12pt
\hbox{hep-th/0209002}
\hbox{MIT-CTP-3300}
\hbox{HUTP-02/A042}
}}
 {\vbox{
 \centerline{Operator Mixing and the BMN Correspondence}
 }}

\centerline{
Neil R. Constable,${}^{1}$
Daniel Z. Freedman,${}^{2}$ Matthew Headrick,${}^{3}$ and Shiraz Minwalla${}^{3}$
}

\bigskip
\centerline{${}^{2}$~Department of Mathematics,}
\centerline{${}^{1,2}$~Center for Theoretical Physics and Laboratory for Nuclear Science}
\centerline{Massachusetts Institute of Technology}
\centerline{Cambridge, Ma. 02139}
\centerline{\tt constabl@lns.mit.edu, dzf@math.mit.edu}
\centerline{}
\centerline{${}^{3}$~Jefferson Physical Laboratory }
\centerline{Department of Physics}
\centerline{Harvard University}
\centerline{Cambridge, MA 02138}
\centerline{\tt headrick@pascal.harvard.edu, minwalla@fas.harvard.edu}

\vskip .3in
\centerline{\bf Abstract}
In this note we update the discussion of the BMN correspondence and
string interactions in hep-th/0205089 to incorporate the effects of
operator mixing. 
We diagonalize the matrix of two point functions of single and double trace
operators, and compute the eigen-operators and their anomalous dimensions to order 
$g_2^2 \lambda'$. Operators in different R symmetry multiplets
remain degenerate at this order; we propose this is a consequence of 
supersymmetry. We also calculate the corresponding energy 
shifts in string theory, and find a discrepancy with field theory results,
indicating possible new effects in light-cone string field theory.

\smallskip

\Date{}

\listtoc
\writetoc

\newsec{Introduction}

Several aspects of the correspondence \bmn\ between
operators of ${\cal N}=4$ $SU(N)$ super-Yang-Mills theory at large
R-charge $J$ and type IIB string theory in a pp-wave background
geometry \refs{\hull,\metsaev} have been investigated in the recent
literature. The map between field theory operators and free string theory
in the pp-wave background  was established by \bmn. In our paper \G\ we 
attempted to extend this map to a correspondence between Yang-Mills 
correlators and interactions in the pp-wave background. In this note we 
will update and correct the discussion presented in \G.  

Field theory in the BMN
limit, $N \rightarrow \infty$, $J \rightarrow \infty$ with
$J^2/N$ fixed, appears to be governed by two parameters
\refs{\bmn,\G,\germone,\BerensteinSA}: an effective gauge coupling $\lambda' =
g^2_{\rm YM}N/J^2$, and an effective genus counting parameter $g_2 =
J^2/N$. These quantities can be expressed in terms of the string
scale $\alpha'$ and coupling $g_{\rm s}$, light-cone momentum $p^+$,
and transverse string mass $\mu$ as
\eqn\params{ 
\lambda^{\prime}= \frac{1}{(\mu p^+ \alpha')^2},
\qquad
g_2 = 4\pi g_{\rm s}(\mu p^+ \alpha')^2.
}

In \G\ we proposed that free three-point
functions of the BMN operators 
\eqn\obmn{O^J_n \equiv 
\frac1{\sqrt{JN^{J+2}}}\sum_{l=0}^Je^{2\pi inl/J}\Tr(\phi Z^l\psi Z^{J-l})}
are related to matrix elements of the string field theory light-cone 
Hamiltonian according to 
\eqn\vertex{
\<i|H_{\rm int}(|j\>\otimes |k\>) = \mu g_2
    (\Delta_i-\Delta_j-\Delta_k) C_{ijk},
}
where $|i\>,|j\>,|k\>$ are the dual string states and 
$\<O_i \bar O_j \bar O_k \> = g_2 C_{ijk}$ 
(we have factored out the trivial 
space time dependence). This proposal was intended to describe
string transitions in which the initial
and final states have the same number of excitations, so that
$\Delta_i-\Delta_j-\Delta_k$ is of order $\lambda^{\prime}$. The proposal
applies in the limit of large $\mu$,
corresponding to weak coupling in field theory.
The RHS of \vertex\ was computed in \G\ by a direct evaluation of the 
free three-point coupling $C_{ijk}$. Subsequently, the 
LHS of \vertex\ was obtained in \svt\ (see also \refs{\huang,\lee,\park}) 
as the large $\mu$ limit of the string field theory interaction $H_{\rm int}$
in the pp-wave background \svo. Results of \svt\ 
confirm the proposal \vertex, which thus appears to be on firm ground.

We claimed in \G\ that the proposal
\vertex\ obeys a nontrivial consistency check, which we now review.
Utilizing Hamiltonian matrix elements from \vertex, and second-order non-degenerate perturbation theory, we computed the one loop mass renormalization 
of excited pp-wave string states. We then compared the result to an explicit
computation of the order $\lambda'g_2^2$ correction to the 
anomalous dimension of the corresponding BMN operator. In \G\ we
reported agreement between these two computations. 

In this note we will point out that this agreement is in fact
spurious.\foot{In Appendix D we show that the relation referred to in \G\
as the ``unitarity check'' (with the sign corrected) in fact follows purely
from field theory considerations, without making use of the BMN duality.}
Firstly our result for the anomalous dimension computed in our
paper had the wrong sign (this was pointed out to us by the authors of
\germone). Secondly our computation was incomplete, in that it did not take
into account mixing between single- and double-trace BMN operators. As we
will describe in this paper, this mixing alters the result for the
anomalous dimension at the order under consideration. In this note we will
present a corrected result for the anomalous dimension of BMN operators at
order $\lambda'g_2^2$.\foot{Operator mixing was first suggested in the BMN
context in \bianchi. A preliminary version of the present
work was presented by one of us (S.M.) at the conference
Strings 2002. As we were drafting this note, the paper \germtwo\ 
was submitted to the archive with similar methodology and
results, for a larger class of operators. One of us (D.Z.F.) acknowledges
useful discussions with G. Semenoff at the Aspen Center for Physics.
Mixing is also discussed in the very recent paper \gross\ .}

Surprisingly, this corrected anomalous scaling dimension
does not match the result of the analogous perturbative calculation on
the string theory side, as we show in section 3. The discrepancy could be
due to presence of a quadratic contact term of order $g_2^2$ in
the string field theory Hamiltonian. This is an important 
issue.

\newsec{Anomalous dimensions of BMN eigen-operators}

The BMN limit preserves an $SO(4)$ R-symmetry group. The complex impurity
fields $\phi$ and $\psi$ transform in the fundamental representation of an
$SU(2)$ subgroup, and operators containing two such impurities transform in
the $\bf 2\otimes2=1\oplus3$ representation. Our calculations thus split into
independent sectors for the two representations.\foot{We thank H. Verlinde
and L. Motl for pointing this out to us.} For example, if we form
the linear combinations
\eqn\glcpm{
O^{\pm J}_n = {1\over \sqrt{2}}\left(O^J_n \pm O^J_{-n}\right)
}
of the operators \obmn, then $O^{+J}_n$ is a member of a triplet,
while $O^{-J}_n$ is a singlet. In this section we will determine, to order
$\lambda' g_2^2$, the anomalous dimension of the eigen-operators
$\tilde{O}_n^{\pm J}$ that reduce to $O^{\pm J}_n$ at $g_2=0$. A priori the
results could have been different for the two representations.
We find instead that the degeneracy persists to this order in
$g_2$; at the end of this subsection we will argue that this degeneracy 
is a consequence of supersymmetry. 

The mixing problem requires the diagonalization of the matrix of
two-point functions $\langle O_{(1)} \bar O_{(1)} \rangle$, $\langle
O_{(1)} \bar O_{(2)} \rangle$, $\langle O_{(2)} \bar O_{(2)} \rangle$ of
single- and double-trace
operators including free and order $\lambda'$ terms.
The mixed two-point functions are of order $g_2$. To
obtain the eigen-operators to this order we need only the order
$g_2^0$ parts of  $\langle O_{(1)} O_{(1)} \rangle$ and $\langle O_{(2)}
O_{(2)}\rangle$. The eigen-operators are obtained in subsection 2.1, in
a treatment which emphasizes the relation of two- and three-point
functions and the issue of correct conformal behavior. In
subsection 2.2 we go on to find the order $\lambda' g_2^2$
correction to the anomalous dimension of the eigen-operators
$\tilde{O}_n^{\pm J}$ which requires the known \refs{\germone,\G} 
order $g_2^2$ parts of $\langle O_{(1)} O_{(1)} \rangle$. As discussed in
Appendix E, in these calculations we make the assumption that the
single-trace operators do not mix with degenerate triple-trace operators,
which is not yet justified by calculation of the relevant two-point
functions. However, the mode $n=1$ is not degenerate with any multi-trace
operators, so results for $n=1$ are safe from this danger.

\subsec{Two-point functions, three-point functions, and mixing}

The two-point functions between single-trace BMN operators $O^J_n$ was
computed to lowest order in $g_2$ in \bmn: 
\eqn\bmntp{
\langle O^{\pm J}_n \bar{O}^{\pm J}_m\rangle =
\delta_{nm}\left(1 -\l' m^2 \ln(x^2\Lambda^2)\right),
\qquad
\langle O^{+ J}_n \bar{O}^{- J}_m\rangle=0.
}
In order to establish some notation that will be useful below we may
express two-point functions between BMN operators of charge $J_1 < J$ as
\eqn\bmntpm{
\langle O^{\pm J_1}_n \bar{O}^{\pm J_1}_m \rangle =
\delta_{nm}
\left(1 -\l' k^2 \ln(x^2\Lambda^2)\right)
}
where $k=Jm/J_1=m/s$, and we have defined $s=J_1/J$ (this was denoted $y$
in \G).

We now turn to the construction of the matrix of two-point functions between 
single-trace and double-trace BMN operators. We will find it convenient 
to first compute, to order $\lambda' g_2$, the three-point functions of three 
single-trace BMN operators located at distinct spatial points; the requisite
two-point functions may then be obtained allowing two of the insertions 
in the three-point function to approach each other.

The Feynman graphs that contribute to the three-point function between
$O^J_n$, $O^{J_1}_m$, and $O^{J-J_1}\equiv
\Tr(Z^{J-J_1})/\sqrt{(J-J_1)N^{J-J_1}}$ at order $\lambda'$ are of two
kinds, with different space time dependence. This gives the following
general structure for the three-point function:
\eqn\thrept{\langle O^J_n(x_1) \bar{O}^{J_1}_{m}(x_2) 
\bar{O}^{J-J_1}(x_3) \rangle = g_2
C_{nms}
\left[1-\lp \left( a_{nk} \ln(x_{12}\Lambda)^2+ b_{nk} 
\ln(\frac{x_{13}x_{12}
\Lambda}{x_{23}})\right)
\right].}
Here $C_{nms}$ is the free three-point function between the relevant
operators which was computed in \refs{\G},
\eqn\frtp{
C_{nms} = \sqrt{\frac{1-s}{Js}}\frac{\sin^2(\pi ns)}{\pi^2(n-k)^2}
.}
Evaluation of the Feynman diagrams gives the values
\eqn\anb{
a_{nk}= k^2, \qquad b_{nk}=n(n-k).
}
but we find it useful to regard $a_{nk}$ and $b_{nk}$ as unspecified
parameters in much of this section.

Note that \thrept, with the values of $a_{nk}$ and $b_{nk}$ given in 
\anb\ {\it does not} take the form dictated by conformal invariance for
the three-point functions of three operators of anomalous dimension $n^2 \l'$, 
$k^2 \l'$, and 0 respectively. This already indicates that $O_n^J$ does not 
have well defined scaling dimension; we will see this in more detail below.

Transforming \thrept\ to the $O^{\pm J}_n$ basis, and using the properties 
$a_{nk}=a_{-n,-k}$, $b_{nk}=b_{-n,-k}$, we can rewrite the above
three-point function as
\eqn\threeptpm{
\langle O^{\pm J}_n(x_1) \bar{O}^{\pm J_1}_{m}(x_2) 
\bar{O}^{J-J_1}(x_3) \rangle = 
g_2 C^{\pm}_{nms}
\left[1-\lp \left( a^{\pm}_{nk} \ln(x_{12}\Lambda)^2+ 
b^{\pm}_{nk}  
\ln(\frac{x_{13}x_{12}
\Lambda}{x_{23}})\right)
\right]
}
and
\eqn\threptppm{\langle O^{\pm J}_n(x_1) \bar{O}^{\mp J_1}_{m}(x_2) 
\bar{O}^{J-J_1}(x_3) \rangle = 0}
where 
\eqn\definitions{\eqalign{
& C^{+}_{nms}=C_{nms} + C_{-n,m,s} = 
H_{nks} {2(n^2+k^2) \over (n^2 -k^2)^2}, \cr
& C^{-}_{nms}=C_{nms}-C_{-n,m,s} =
H_{nks}\frac{4nk}{(n^2-k^2)^2}, \cr
&H_{nks}=\sqrt{{1-s}\over J s} {\sin^2(n \pi s) \over \pi^2}.}
}
When the values in \anb\ are used
\eqn\anbplus{\eqalign{
&a^\pm_{nk}=k^2, \cr
&b^{+}_{nk}={n^2 (n^2-k^2) \over n^2+k^2}, ~~~
b^{-}_{nk}={n^2-k^2 \over 2}}}

We now turn to the construction of two-point functions between double-trace
and single-trace operators. 
In Appendix A we demonstrate that the two-point function
\eqn\twotwo{\langle O^J_n(0)
:\bar{O}^{J_1}_{m}\bar{O}^{J-J_1}:(x) \rangle \equiv \langle O^J_n(0)
\bar{O}^{J}_{ms}(x) \rangle}
may be obtained from \thrept\ by the replacement $x_{12}=x_{13}\rightarrow x,~~ x_{23}\rightarrow 
\frac{1}{\Lambda}$ so that
\eqn\twopt{\langle O^J_n(0)
\bar{O}^{J}_{ms}(x) \rangle = g_2
C_{nms}
\left(1-\lp \ln(x\Lambda)^2(a_{nk}+b_{nk}) \right).}

The final two-point function that we need  involves two double-trace
operators. It is straightforward to show that the result up to order $g_2$
is
\eqn\ddouble{
\langle
:O^{J_1}_{m}O^{J-J_1}:(0):\bar{O}^{J_1}_{n}\bar{O}^{J-J_1}:(x)\rangle = 
\delta_{mn}\left(1-\lambda^{\prime}k^2\ln(x^2\Lambda^2)\right).
}
The complete matrix of two-point functions can now be assembled from
\bmntp, \twopt\ and \ddouble. If we introduce the index $A = i,j,m,n$
where $m,n$ index the single-trace operators $O^{J}_n$ and $i,j$ index the
double-trace operators $O^{J}_{ms}$ so that $i\equiv (m,s)$ then the matrix
of two-point functions found in this section can be summarized as
\eqn\combine{
\<O_A\bar{O}_B\>=g_{AB}-\lambda^{\prime}h_{AB}\ln{x^2\Lambda^2}.
}
Here $g_{AB}$ gives an inner product on the space of operators (in fact it
is just the Hilbert space inner product, according to the state-operator
mapping), while $h_{AB}$ is the matrix of anomalous
dimensions (obtained by applying the dilatation operator $D=x^{\nu}\partial_{\nu}$). 
See Appendix C for a schematic discussion of the
diagonalization of $\langle O_A \bar{O}_{B} \rangle$.

\subsec{Construction of eigen-operators}

In order to find eigen-operators we must now diagonalize the matrix of 
two-point functions. The eigen-operators $\tilde{O}^{\pm J}_n$ and 
$\tilde{O}^J_{ms}$ must, by definition, reduce to $O^{\pm J}_n$ and 
$:{O}^{J_1}_{m}O^{J-J_1}:$ respectively at $g_2=0$. These operators 
take the form
\eqn\diaoguess{\eqalign{
\tilde{O}^{\pm J}_n &=
O^{\pm J}_n + g_2\sum_{m=0}^{\infty}\sum_{J_1=0}^{J} C^{\pm}_{nms}
M_{nm} :O^{\pm J_1}_m O^{\pm J-J_1}:+\cdots \cr
\tilde{O}^{\pm J}_{ms} &=
{}:{O}^{\pm J_1}_{m}O^{\pm J-J_1}:-g_2\sum_{m=0}^{\infty}C^{\pm}_{nms}
N_{nm} O^{\pm J}_n +\cdots
}}
where $M_{nk}$ and $N_{nk}$ are coefficients to be determined.\foot{The transformation to 
the $\pm$ basis implies that the summation symbol in \diaoguess\
and other formulas in this basis should be defined as $\sum_{m=0}^{\infty} \equiv
\frac{1}{2}\sum_{m=-\infty}^{\infty}$.}
The two-point function between $\tilde{O}^{\pm J}_n$ and 
$\tilde{O}^{\pm J}_{ms}$ is easily computed using the results of the previous 
subsection
\eqn\tpsd{\eqalign{
(4 \pi^2 x^2)^{J+2} &
\langle \tilde{O}^{\pm J}_n(0) \bar{\tilde O}{}^{\pm J}_{ms}(x) \rangle \cr 
& = g_2 C^{\pm}_{nms}\left[ \left(1+M_{nm}-N_{nm} \right) -
\lambda' \ln(\Lambda x)^2
\left(a^{\pm}_{nk}+b^{\pm}_{nk} +M_{nm}k^2-N_{nm}n^2 \right)
\right] .}}
As $\tilde{O}^{\pm J}_n$ and $\tilde{O}^{J}_{ms}$ are eigen-operators in a
conformal field theory, \tpsd\ must vanish identically; this yields
the set of simultaneous equations 
\eqn\sime{\eqalign{
1+M_{nm}-N_{nm}& = 0 \cr
a^{\pm}_{nk}+b^{\pm}_{nk} +M_{nm}k^2-N_{nm}n^2 &= 0
}}
which may easily be solved for $M_{nm}$ and $N_{nm}$. Carrying out the 
algebra we find
\eqn\diao{\eqalign{
\tilde{O}^{\pm J}_n &= O^{\pm J}_n +g_2
\sum_{m=0}^\infty \sum_{J_1=0}^J C^{\pm}_{nms}
{a^{\pm}_{nk}+b^{\pm}_{nk}-n^2 \over 
n^2-k^2}O^{\pm J_1}_m O^{J-J_1}+\cdots \cr
\tilde{O}^{\pm J}_{ms}&={:O^{\pm J_1}_{m}O^{J-J_1}:}-g_2\sum_{m=0}^{\infty}
C^{\pm}_{nms}
{a^{\pm}_{nk}+b^{\pm}_{nk}-k^2 \over n^2-k^2}O^{\pm J}_n +\cdots.
}}

\subsec{Consistency conditions from three-point functions}

Having determined the BMN eigen-operators to first order in $g_2 \lambda'$, 
we now turn to the determination of their three-point functions. We will
demonstrate that the three-point functions of these operators do indeed
take the form required by conformal invariance. We find that this is true
for any value of $b_{nk}$ provided the coefficient $a_{nk}=k^2$. This is
consistent with \anb\ and provides a check on our algebra.

Combining \bmntp, \bmntpm, \thrept, and \diao\ (and using the fact that
double trace correlators factorize to lowest order in $g_2$) we find 
\eqn\threptg{\eqalign{&\langle \tilde{O}^{\pm J}_n(x_1) 
\bar{\tilde{O}}^{\pm J_1}_{m}(x_2) 
\bar{O}^{J-J_1})(x_3) \rangle = g_2
C^{\pm J}_{nms}{a^{\pm}_{nk}+b^{\pm}_{nk}-k^2 \over n^2-k^2} \cr
&\times\left(1-
\lp\left[\ln|x_{12}\Lambda| {2n^2k^2-b^{\pm}_{nk} (k^2+n^2)-
2a^{\pm}_{nk} n^2 \over k^2-a^{\pm}_{nk} -b^{\pm}_{nk}}
+\ln|\frac{x_{13}x_{12}\Lambda}{x_{23}}|
b^{\pm}_{nk} {k^2-n^2 \over k^2-a^{\pm}_{nk} -b^{\pm}_{nk}} 
\right]\right).}}
\threptg\ takes the standard CFT form for a three-point function between
operators  of  anomalous  dimension  $n^2$,  $k^2$,  and  $0$  respectively,
provided that 
\eqn\cond{\eqalign{
{2n^2k^2-b^{\pm}_{nk} (k^2+n^2)-2a^{\pm}_{nk} 
n^2 \over k^2-a^{\pm}_{nk} -b^{\pm}_{nk}} &= k^2+n^2 \cr 
b^{\pm}_{nk} {k^2-n^2 \over k^2-a^{\pm}_{nk} -b^{\pm}_{nk}}
&= k^2.
}}
It is easily verified that the two equations \cond\ are not independent;
they are both satisfied for any value of $b_{nk}$ if and only if
\eqn\deta{
a^{\pm}_{nk}=k^2,
}
as promised at the beginning of this subsection.
 
Notice that the three-point coupling between three normalized eigen-operators 
involves the coefficient  $$\tilde{C}^{\pm}_{nms}=C^{\pm}_{nms}{b^{\pm}_{nk} \over n^2-k^2}$$
which is distinct from the free three-point coupling $C^{\pm}_{nms}$
even at lowest order in $\lambda'$! Even though this modification is 
independent of $\lambda'$, it is clearly a quantum effect as 
the modified three-point function depends on $b_{nk}$, the coefficient of 
the order $\lambda'$ term in \threeptpm.

\subsec{Anomalous dimensions of BMN eigen-operators}

It is now a simple matter to compute the two-point function of the operator 
\diao\ and thereby determine its anomalous dimension to order
$g_2^2\lambda'$.\foot{We do not need to know the form of the operator to
order $g_2^2$ in order to perform this computation. The only order $g_2^2$
correction to this operator that can contribute to its two-point function
at order $g_2^2$ is a piece proportional to $O^J_n$;  such an addition
can  be absorbed into  a normalization of the operator at this order, and
so does not contribute to its scaling dimension.} The two-point
functions of single-trace BMN operators with double-trace operators were
presented above while the complete order $\lambda'g_2^2$ two-point functions of
single-trace BMN operators were presented in \refs{\germone, \G} and we
reproduce here (with different relative signs than \refs{\G}) the relevant
equations,
\eqn\oldtwopt{
\<O_n^J(0)\bar{O}_m^J\> = \left(\delta_{nm}+g_2^2A_{nm}\right)\left
(1-(n^2-nm+m^2)\lambda^{\prime}\ln{x^2\Lambda^2}\right)
-\frac{g_2^2\lambda^{\prime}}{4\pi^2}B_{mn}\ln{x^2\Lambda^2}
}
where the matrices $A_{mn}$ and $B_{mn}$ are given by
\eqn\coefa{
A_{nm} = \cases{
\frac1{24} & if $m=n=0$ \cr
0, & if $ m=0, n\neq0 $ { or } $n=0\,, m\neq0 $ \cr
\frac1{60} - \frac1{6u^2} + \frac7{u^4} & if $m=n\neq0 $ \cr
\frac1{4u^2}\left(\frac13+\frac{35}{2u^2}\right)  & if $m=-n\neq0 $ \cr
\frac1{(u-v)^2}
\left(\frac13+\frac4{v^2}+\frac4{u^2}-\frac6{uv}-\frac2{(u-v)^2}\right) &
{all other cases} \cr}}
\eqn\coefb{
B_{nm} = \cases{
0 & n=0 if $m=0$ \cr
\frac13+\frac{10}{u^2} & if $n=m\neq0 $ \cr
-\frac{15}{2u^2} & if $n=-m\neq0 $ \cr
\frac6{uv}+\frac2{(u-v)^2} & {all other cases,} \cr }}
and
$$u = 2\pi m,\qquad v = 2\pi n .$$

As explained in Appendix C, the anomalous dimension of the $\tilde{O}^{\pm
J}_n$ can be readily computed from the diagonal two-point functions
$\langle \tilde{O}^{\pm}_n \bar{\tilde{O}}^{\pm}_n \rangle$. The required
algebra is straightforward using the ingredients  \twopt, \diao,
\oldtwopt, \coefa, \coefb. The result is
\eqn\anomdimadd{
{\Gamma^{\pm}_n \over \l' g_2^2} = 
\mp 2 n^2 A_{n, -n} +{1 \over 4 \pi^2}(B_{nn}\pm B_{n,-n})
- \sum_{m=0}^{\infty}\sum_{J_1=0}^{J} (C^{\pm}_{nms})^2{(b^{\pm}_{nk}-n^2+k^2)^2 \over n^2-k^2}.
}
where we use the value for $b^{\pm}_{nk}$ given in \anbplus\ .
Using the identities in Appendix D this may be re-written as
\eqn\result{
\Gamma_n^\pm = 
g_2^2\lambda^{\prime}n^2A_{n-n} =
{g_2^2\lambda^{\prime}\over 4 \pi^2}\left({1\over 12}+{35 \over 32
\pi^2n^2}\right)
}
The full anomalous dimension of this operator (including the planar contribution) 
is
\eqn\andim{\Delta_n^{\pm}-J-2 = 
\lambda^{\prime}n^2\left(1+g_2^2A_{n,-n}\right) = 
\lambda'\left[
n^2 + \frac{g_2^2}{4\pi^2}\left(\frac{1}{12} +\frac{35}{32\pi^2n^2}\right)
\right].
}
This is the principal result of this note. 

It is striking that the degeneracy between the even and odd operators 
(obvious at $g_2=0$) persists to order $g_2^2$. It appears that this
degeneracy is a consequence of supersymmetry\foot{This explanation was 
suggested to us by J. Maldacena and M. Van Raamsdonk in response to an
earlier version of this paper.}; all two-impurity BMN operators lie in a 
single representation of the pp-wave supersymmetry algebra and so are 
guaranteed to have equal anomalous dimensions. Note that 16 of the 32 
supercharges of the $\CN=4$ $d=4$ superconformal algebra commute with 
$p^-={\mu \over 2}(\Delta-J)$; further all supercharges commute with 
$p^+$ which is 
a central element of this algebra in the pp-wave limit. Consequently, 
states in long representations of the pp-wave superalgebra 
appear in multiplets whose minimum degeneracy is $2^{16 /2}=16
\times 16$. But this is precisely the number of two-impurity BMN
operators; it is natural to conjecture that the two-impurity BMN operators 
may be related to one another by the action of the 16 supercharges that 
commute with $p^-$; their degeneracy is thus a consequence of
supersymmetry. Special examples of such relations have been worked out
directly in the gauge theory in \refs{\sz,\GursoyYY}. We hope to return to
this issue in more detail in future work.

\newsec{String field theory revisited}

In this section we will use quantum mechanical perturbation theory to 
compute the energy shifts of a particular set of excited string states
in the pp-wave background. This section is almost a direct transcription 
of Section 5 of \G, now adapted to the $O^{\pm J}_n$ basis. 

Utilizing \vertex\ and second order non-degenerate perturbation theory 
(see section 5 of \G) we find, 
\eqn\evench{\eqalign{
\frac1\mu\sum_{i}\frac{|\<n|H_{\rm int}|i\>|^2}{E_n-E_i}   
= g_2^2\l'\sum_i(n^2-k^2)C_{ni}^2 = -\frac{g_2^2\l'}{4\pi^2}B_{nn} = 
-\frac{g_2^2\l'}{4\pi^2}\left(\frac13+\frac{10}{v^2}\right) 
}}
and 
\eqn\oddch{\eqalign{
\frac1\mu\sum_{i}\frac{\<n|H_{\rm int}|i\>\<i|H_{\rm int}|-n\>}{E_n-E_i}  = 
g_2^2\l'\sum_i(n^2-k^2)C_{ni}C_{-n,i} = 
-\frac{g_2^2\l'}{4\pi^2}B_{n,-n} = 
\frac{g_2^2\l'}{4\pi^2}\frac{15}{2v^2}.
}}
where we have used $i$ as a collective index for $(m,s)$ as well as for the
states in which the $\phi$ and $\psi$ excitations are on different strings,
corresponding to the BPS operator $:O_\phi^{J_1}O_\psi^{J-J_1}:$.
Unlike in the Yang-Mills calculation of the previous section, the
degeneracy between the $+$ and $-$ sectors of the string theory is thus
broken: the correct zeroth-order eigenvectors are
\eqn\evod{
|n,\pm\> \equiv \frac{1}{\sqrt{2}}\left(|n\>\pm|-n\>\right),
}
and their second order energy shifts are
\eqn\perten{\eqalign{
\frac1\mu\sum_{i}\frac{|\<n,+|H_{\rm int}|i,+\>|^2}{E_{n,+}-E_{i,+}} = 
-\frac{g_2^2\l'}{4\pi^2}(B_{nn}+B_{n,-n}) = 
-\frac{g_2^2\l'}{4\pi^2}\left(\frac13+\frac5{2v^2}\right)  \cr
\frac1\mu\sum_{i}\frac{|\<n,-|H_{\rm int}|i,-\>|^2}{E_{n,-}-E_{i,-}} 
= -\frac{g_2^2\l'}{4\pi^2}(B_{nn}-B_{n,-n}) = 
-\frac{g_2^2\l'}{4\pi^2}\left(\frac13+\frac{35}{2v^2}\right).
}}
Sadly, neither of these energy shifts agrees with the anomalous dimension 
computed in the previous section.

It is possible that this disagreement is resolved by the presence of an 
explicit quadratic contact term in the string field theory Hamiltonian. 
In order to restore the degeneracy of string states and resolve the
discrepancy, this contact term would have to have matrix elements 
\eqn\conteig{\eqalign{
\frac1\mu\<n,+|V_2|n,+\> = 
\frac{g_2^2\l'}{4\pi^2}\left(\frac5{12}+\frac{55}{8v^2}\right) \cr
\frac1\mu\<n,-|V_2|n,-\> =
\frac{g_2^2\l'}{4\pi^2}\left(\frac5{12}+\frac{175}{8v^2}\right).
}}
Since these expressions are positive it is at least possible that they
come from a contact term in the string field theory Hamiltonian, which
would be determined by the anti-commutator of two supersymmetry
generators \refs{\GreensiteHM,\GreenTC,\VerlindeIG}. It would be very
interesting to compare \conteig\ with a direct computation of contact terms
in string field theory.

It should also be mentioned that the energy shifts could possibly receive
contributions at this order from states with 4 or more impurities, due to
enhanced matrix elements between states with different numbers of
impurities, as reported in \svt. Note, however, that this effect could not
alone cure the apparent discrepancy between string theory and gauge theory,
since the contributions, if any, due to such massive intermediate states
would be negative.

\bigskip

\noindent
{\bf Acknowledgements}
\bigskip
We would like to thank V. Khoze, C. Kristjansen, L. Motl, J. Plefka,
A. Postnikov, G. Semenoff, W. Skiba, M. Spradlin, M. Staudacher,
A. Volovich and especially J. Maldacena, M. Van Raamsdonk and 
H. Verlinde for useful discussions and correspondence. 
S. M. would like to thank the Tata Institute for
Fundamental Research, Mumbai for hospitality while this work was being 
completed. S.M. and D.F.Z would like to thank the organizers of the 
Amsterdam Summer workshop for hospitality while this work was in progress.
 N.R.C. and D.Z.F. are supported by the NSF under PHY
00-96515 and the DOE under DF-FC02-94ER40818. N.R.C. is also supported by
NSERC of Canada. M.H. is supported in part by the DOE under grant
DE-FG01-91ER40654 and in part by the Packard Foundation. S.M. is supported
by a Harvard Junior Fellowship and by the DOE grant DE-FG01-91ER40654.

\appendix{A}{The space-time structure of two- and three-point functions}

In this appendix we will justify the space time dependence of the order
$\lambda'$
terms in the equation \thrept\ for the three-point function
$\langle O^J_n(x_1) \bar{O}^{J_1}_{m_1}(x_2)
\bar{O}^{J_2}(x_3) \rangle$, and discuss its relation to mixed
two-point functions as in \twopt. There are two distinct
structures coming from the quartic interactions in contributing
Feynman diagrams. The first structure occurs in diagrams in
which one pair of lines from the operator $O^J_n(x_1)$ and one
pair from $\bar{O}^{J_1}_{m_1}(x_2)$ terminate at the interaction
vertex at $z$. This leads to the the same space time integral
which occurs in two-point functions, namely
\eqn\sio{
\int \frac{d^4z}{z^4(z-x_{12})^4} = 
2\pi^2 \frac{\ln(x_{12}^2\Lambda^2)}{x_{12}^4}.
}
The second structure occurs in diagrams
where lines from all three operators are connected to the interaction
vertex. This leads to the integral
\eqn\sit{
\int\frac{d^4z}{z^4(x_{12}-z)^2(x_{13}-z)^2} =
\pi^2 \ln\left(\frac{x_{12}^2x_{13}^2\Lambda^2}{x_{23}^2}\right)
}
This integral was evaluated using
dimensional regularization in \chu\ and with differential
regularization \diffreg\ in our work. The integrals above determine the
general structure in \thrept, and the specific values of the
parameters $a_{nk},b_{nk}$ which come from our summation of the
contributing diagrams are given in \anb.

The two-point function $\langle O^J_n(x_1) :\bar{O}^{J_1}_{m_1}
\bar{O}^{J_2}:(x_2)\rangle $ can be obtained from the three-point
function in the OPE limit $x_3 \rightarrow x_2$. See \chutoo. It
can also be found simply by setting $x_{13}=x_{12}$ and
$x_{23}= 1/\Lambda$, since \sit\ reduces to \sio\ if this
is done. Two-point functions were also obtained directly from
summation of Feynman diagrams in our work.

\appendix{B}{Absence of mixing with BPS double-trace operators}

In \refs{\G}---see also \refs{\huangtwo,\chu}---non-vanishing three-point
functions $\langle O^J_n \bar{O}^{J_1}_\phi \bar{O}^{J-J_1}_\psi \rangle $
were computed, where $O^{J}_\phi = \Tr(\phi Z^J)/\sqrt{N^{J+1}}$ and
$O^{J}_\psi = \tr(\psi Z^J)/\sqrt{N^{J+1}}$ are BPS operators. The
corresponding dual two-string state contributed to the sum over
intermediate states in the consistency check in \G. One may then wonder why
the double-trace $:\bar{O}^{J_1}_\phi \bar{O}^{J-J_1}_\psi:$ was
not included in the initial ansatz \diao\ for our
mixing calculation, and we now justify its omission.

Let $O^J = \Tr(Z^J)/\sqrt{JN^J}$ denote the chiral primary operator. It is
quite easy to verify that the linear combination
\eqn\susyop{
O^J_{\rm susy} \equiv 
\left(\sqrt{\frac{J_1-1}{J_2+1}} :O^{J_1-1}_0O^{J_2+1}: + :O^{J_1}_\phi
O^{J_2}_\psi:\right)
+ \left( J_1 \leftrightarrow J_2\right)
}
is an $SU(4)$ descendant of the chiral primary
$:\Tr(Z^{J_1+1})\Tr(Z^{J_2+1}):$. Thus the operator
$O^J_{\rm susy}$ decouples from $O^J_n$ and from
$:O^{J_1}_{m}O^{J_2}:$ for $n,m \ne
0$. The decoupling can easily be observed; the
appropriate combination of mixed two-point functions which can be
obtained from the three-point functions given in (3.10) and
(3.11) of \refs{\G} vanish.\foot{To correct an error the right side of
(3.11) in \G\ should be multiplied by $-1$.}
The decoupling means that we can simply
omit the operators  $:\bar{O}^{J_1}_\phi \bar{O}^{J_2}_\psi:$
from the mixing calculation, if we keep $:O^{J_1}_0 O^{J_2}:$.
However, one may also observe directly from the solution for
$\tilde{O}^J_n$ in \diao\ that $:O^{J_1}_0 O^{J_2}:$ also
decouples.

\appendix{C}{Diagonalization procedure}

In this appendix we provide an overview of the diagonalization
procedure used to
obtain the eigen-operators in \diao\ and their scale dimension. 

The matrix of two point functions is given in \combine\. 
The problem is to diagonalize the matrix of 
anomalous scaling dimensions  $h_{AB}$ relative to the 
inner product $g_{AB}$. Both $h_{AB}$
and $g_{AB}$ are functions
of $g_2$ and we will perform our analysis up to and including
contributions at order $g_2^2$. To explain the notion of relative
diagonalization, we note that the goal is to find a set of operators,
\eqn\OA{\tilde{O}_A = O_B S^B{}_A}
and numbers $\gamma_A$ such that
\eqn\diag{
(4\pi^2x^2)^{J+2}\left\<\tilde{O}_A(0)\bar{\tilde{O}}_B(x)\right\> =
\delta_{AB}(1-\l \gamma_A \ln x^2\Lambda^2).}
Hence the matrix $S$ must satisfy
\eqn\S{S^{\dag} gS=I, \qquad S^{\dag} hS = \gamma}
where $\gamma_{AB} = \gamma_A\delta_{AB}$ is the diagonalized matrix of
anomalous scaling dimensions. Together these imply
\eqn\SS{
S^{-1}g^{-1}hS = \gamma.
}
Thus we must find the eigenvalues and eigenvectors of the matrix
\eqn\mat{
{h^A}_C = g^{AB}h_{BC},
}
where $g^{AB}$ is the inverse matrix of $g_{AB}$.
This diagonalization process is essentially equivalent to
non-degenerate quantum mechanical perturbation theory. We use a
two-step process in which we first find the eigenvectors to order $g_2$,
and then use the two-point function of the eigen-operators to
compute the scale dimension to order $g_2^2$.

To make the method clear consider the toy eigen-value problem
\eqn\toy{
      M^a{}_b V^b_{(c)} = \lambda_{(c)}  V^a_{(c)}.
}
Diagonal elements are given by $M^a{}_a = \rho_{(a)}$, and we wish to
work to first order in all off-diagonal elements. To zero order,
the eigenvalues are  $\lambda_{(c)}= \rho_{(c)}$ and eigenvectors
are   $V^a_{(c)} =\delta^a{}_c.$ It is then trivial to see that the
first order eigenvectors are
\eqn\evects{\eqalign{
V^a_{(c)} &= 1 \,\,\,\,{\rm if}\,\,\, a=c  \cr
V^a_{(c)} &= \frac{M^a{}_c}{\rho_{(c)}- \rho_{(a)}} \,\,\,\,{\rm if} \,\,\,a \ne c
}}
The correction is well-defined if there is no degeneracy. This
toy model is pertinent to our block matrix $h^A{}_B$ because the
non-diagonal corrections to $h^n{}_m$ and $h^{i}{}_{j}$ are of
order $g_2^2$ and can be temporarily ignored. One immediately
finds the (dominantly single-trace) eigenvectors to first order in $g_2$:
\eqn\evecp{
\tilde{O}_n = O_n + \sum_i\frac{O_i{h^i}_n}{n^2-k^2} =
O_n -g_2\sum_i \frac k{n+k}C_{n,i}O_i.
} 
One may also write an analogous expression for the dominantly
double-trace eigenvectors of the structure $\tilde{O}_{i} =
O_{i} +\CO(g_2 O_m)$.  From linear combinations of $O_n$ and
$O_{-n}$ one can easily obtain the eigen-operators in \diao\ 
 (with \anb\ inserted). 

The second step of our procedure is based on the matrix of
two-point functions $\langle \tilde{O}_A \bar{\tilde{O}}_B
\rangle$ (in the $\pm$ basis). Because of the approximate
diagonalization,
off-diagonal entries
 $\langle \tilde{O}_n \bar{\tilde{O}}_i \rangle$ are of order
$g_2^3$ and can be dropped. We can confine our
attention to the upper block of the matrix which is again of the form
\eqn\upbk{
\langle\tilde{O}_n^{\pm}~\tilde{O}_m^{\pm}\rangle =
G_{nm} - \l H_{nm}\ln x^2 .}
(with superscripts omitted on the RHS to avoid clutter). We need
to think about diagonalizing the matrix $H^n_m = G^{np}H_{pm}$, in
which $G_{nm},H_{nm}$ have the structure
\eqn\blah{\eqalign{
G_{nm} &= \delta_{nm} + g_2^2 g^{\prime}_{mn}\cr
H_{nm} &= n^2 \delta_{nm} + g_2^2 h^{\prime}_{nm} }}
The off-diagonal elements of $H^n_m$ are of order
$g_2^2$. There is no degeneracy so these affect eigenvalues
beginning only in order $g_2^4$. Much as in the toy
model above, the anomalous
dimension of the eigen-operator $\tilde{O}_n^{\pm}$ is then simply the
diagonal element
\eqn\blue{
H^n_n = n^2 +g_2^2 (h'_{nn} -n^2g'_{nn}).}
It is this quantity in the $\pm$ channels that is computed in \anomdimadd\ .

\appendix{D}{Deconstruction identities}

In this appendix we note several useful identities which relate the three 
point and two-point coefficients that appear in this paper. We follow the
notation of section 3 in which $i$ is used as a collective index labeling
both types of double-trace operators: $:O_m^{J_1}O^{J-J_1}:$ (for which
$s=J_1/J$ and $k=m/s$) and $:O_\phi^{J_1}O_\psi^{J-J_1}:$ (for which
$s=J_1/J$ and $k=0$). We then have:
\eqn\decon{\eqalign{
\sum_i C_{ni} C_{mi} &= 2A_{nm} \cr
\sum_i k C_{ni} C_{mi} &= (n+m)A_{nm} \cr
\sum_i k^2 C_{ni} C_{mi} &= (n^2+m^2)A_{nm} + {1\over4\pi^2}B_{nm}.
}}
The first of these was pointed out in \huang. The first and last can be
combined to obtain the relation
\eqn\unicheck{
\sum_i(n^2-k^2)C_{ni}^2 = -{1\over4\pi^2}B_{nn},
}
which is the content (up to a sign) of the ``unitarity check'' of \G.

These formulas are not numerical coincidences, nor do they in and of
themselves provide evidence for the BMN correspondence, since they can be
derived purely within the gauge theory. Nonetheless the derivation, which
we will explain in this appendix, naturally calls to mind a picture of
interacting strings, and may in the future be useful in gaining a deeper
understanding of the correspondence.

We imagine the string of $Z$s in the operator $O_n^J$ to be a circle on
which the $\phi$ and $\psi$ impurities are quantum mechanical
particles moving with momentum $n$ and $-n$ respectively---their wave
function is the weight of each possible configuration in the definition of
the operator \obmn. (Morally this point of view is a ``de-second
quantization'' of the worldsheet field theory, which is appropriate in the
large $\mu$ limit where the excitations are very massive
and we are fixing their number and type.) A free torus diagram contributing
to $A_{nm}$ can be represented by a unitary operator $T(s,j_1,j_3)$ which
acts on this 2-particle Hilbert space by dividing the circle (of length 1,
for simplicity) into 4 blocks of lengths $j_1,j_2,j_3,j_4$ (where
$j_2=s-j_1$, $j_4=1-s-j_3$) and transposing blocks 1 and 2 and blocks 3 and
4 (see figure 3 of \G, for example). We thus have
\eqn\aone{
A_{nm} =
{}_\phi\<n|{}_\psi\<{-n}|R|n\>_\phi|{-n}\>_\psi
}
where
\eqn\atwo{
R = 
\frac14\int_0^1ds\int_0^sdj_1\int_0^{1-s}dj_3\,T(s,j_1,j_3)
}
(the integrals count each diagram 4 times, hence the $1/4$). But if we
imagine pinching the circle into two circles of lengths $s$ and $1-s$,
$T(s,j_1,j_3)$ acts by rotating the $s$ circle through an angle $2\pi j_1$
and the $1-s$ circle through an angle $2\pi j_3$. The integration over
these rotation angles will ensure a factorization over physical
intermediate states. Let us see this more explicitly.

In such a pinching, the
2-particle Hilbert space decomposes into a direct sum of four Hilbert
spaces depending on which little circle each impurity is on:
\eqn\athree{
\H_\phi(1)\H_\psi(1) =
\H_\phi(s)\H_\psi(s) \oplus \H_\phi(1-s)\H_\psi(1-s) \oplus
\H_\phi(s)\H_\psi(1-s) \oplus \H_\phi(1-s)\H_\phi(s).
}
The integral over the rotation angle $j_1$ projects out any state with
non-zero total momentum on the $s$ circle, and similarly the integral over
$j_3$ projects out any state with non-zero total momentum on the $1-s$
circle. Hence we have
\eqn\afour{
\int_0^sdj_1\int_0^{1-s}dj_3\,T(s,j_1,j_3) =
s(1-s)\left(
\sum_m P_m(s) + \sum_m P_m(1-s) + P_{0'}(s) + P_{0'}(1-s)
\right),
}
where $P_m(s)$ projects onto the state $|m,s\>_\phi|{-m,s}\>_\psi$
corresponding to the operator $:O_m^{J_1}O^{J-J_1}:$, and $P_{0'}(s)$
projects onto the state $|0,s\>_\phi|0,1-s\>_\psi$ corresponding to the
operator $:O_\phi^{J_1}O_\psi^{J-J_1}:$. Using
\eqn\afive{
C_{ni} = \sqrt{s(1-s)}\cases{
{}_\phi\<n|{}_\psi\<{-n}|m,s\>_\phi|{-m,s}\>_\psi \quad {\rm for }
:O_m^{J_1}O^{J-J_1}: \cr
{}_\phi\<n|{}_\psi\<{-n}|0,s\>_\phi|{0,1-s}\>_\psi \quad {\rm for }
:O_\phi^{J_1}O_\psi^{J-J_1}:
}}
we recover the first identity of \decon.

Inserting $k^2/2$ in the sum means including interaction vertices in the
diagrams. However, the vertex is included only if it is between two lines on
the same small circle. Thus (in the language of \G) all the nearest
neighbor interactions are counted, but only half the semi-nearest neighbor
interactions. The non-nearest neighbor interactions are also under counted
by a factor of 2, but this is a little more complicated: only 1/3 of them
are counted, but overall diagrams with three blocks are over counted by a
factor of 3/2. Hence we have
\eqn\bone{
\frac12\sum_i k^2C_{ni}C_{mi} = 
nmA_{nm} + \frac12(n-m)^2A_{nm} + \frac1{8\pi^2}B_{nm},
}
which is equivalent to the last identity in \decon.

The second identity of \decon\ can also be derived by similar but more
involved reasoning.

\appendix{E}{Degenerate operators and mixing}

The diagonalization process is an application of 
non-degenerate quantum mechanical perturbation theory. A
potential complication occurs because single and double-trace
operators have equal zeroth-order eigenvalues when $n=k$, threatening a
divergence in the calculation of eigenvectors to order $g_2$.
This would signal the breakdown of non-degenerate perturbation theory
and the necessity of using the degenerate theory. As one can see from the
right-hand side of \diao, such a divergence does not occur, thanks to some
rather delicate cancellations. These cancellations have an interesting 
reflection in the dual string theory \G. The amplitude for an excited string 
state corresponding to a 2 impurity BMN operator to decay into two
particles vanishes onshell at lowest order in ${1\over \mu}$; thus these 
excited string states are stable to lowest order in $g_2$.

Let us now consider higher rank $r \ge 3$ multi-trace
operators which we call $O_r$ for short. The two-point functions
$\<\bar O_n O_r\>$ are of order $g_2^{r-1}$. Generically such operators
contribute to eigenvalues beginning in higher order than $g_2^2$. There
could be a complication for triple-trace operators
$O_3={:O_m^{J_1}\Tr(Z^{J_2})\Tr(Z^{J_3}):}$ which are degenerate with
single-traces when $n=m/s$. If the relevant interaction matrix
elements do not vanish, eigen-operators change at zeroth order and
scale dimensions at order $g_2^2$. The vanishing or otherwise 
of these interaction matrix elements is linked to the vanishing or
otherwise of the onshell one-three particle decay amplitudes of string 
field theory.

Degeneracy cannot occur for momentum
$n=1$ since $m \in Z^+$ and $0<J_1<J$. A calculation of
two-point functions $\<\bar O_n O_3\>$
to order $g_2^2$ and $\lambda' g_2^2 $ would be needed to see if the
interaction matrix elements vanish, as they do for
double-traces. The
calculation has not been done yet, so a conservative position is
that present  mixing results are valid for momentum
$n=1$ where degeneracy cannot occur, but may need
modification for $n \ge 2$ if such mixing does occur.

\listrefs

\end